\documentclass{article}

\usepackage{arxiv}

\usepackage[utf8]{inputenc} % allow utf-8 input
\usepackage[T1]{fontenc}    % use 8-bit T1 fonts
\usepackage{hyperref}       % hyperlinks
\usepackage{url}            % simple URL typesetting
\usepackage{booktabs}       % professional-quality tables
\usepackage{amsfonts}       % blackboard math symbols
\usepackage{nicefrac}       % compact symbols for 1/2, etc.
\usepackage{microtype}      % microtypography
\usepackage{lipsum}
\usepackage{graphicx}

\usepackage{algorithm}
\usepackage{algorithmic}
\usepackage{svg}
\usepackage{subfig}
\usepackage{xcolor}
\usepackage{threeparttable}

\usepackage{amsmath}
\usepackage{amssymb}

\graphicspath{ {./images/} }

\title{AST: Adaptive, Seamless, and Training-Free Precise Speech Editing}

\author{
 Sihan Lv$^{1}$\\
  \texttt{shlv@zju.edu.cn} \\
   \And
 Yechen Jin$^{1}$\\
  \texttt{12551012@zju.edu.cn} \\
   \And
 Zhen Li$^{1,3}$\\
  \texttt{lizhen\_cast@zju.edu.cn} \\
   \And
 Jintao Chen$^{1,2}$\\
  \texttt{chenjintao@zju.edu.cn} \\
   \And
 Jinshan Zhang$^{1,2}$\\
  \texttt{zhangjinshan@zju.edu.cn} \\
   \And
 Ying Li$^{1,2,4}$\\
  \texttt{cnliying@zju.edu.cn} \\
   \And
 Jianwei Yin$^{1,2,4}$\\
  \texttt{zjuyjw@cs.zju.edu.cn} \\
   \And
 Meng Xi$^{1,2,4,*}$\\
  \texttt{ximeng@zju.edu.cn} 
}

\renewcommand{\today}{
$^1$ School of Software Technology, Zhejiang University, Hangzhou, China\\
$^2$ Innovation and Management Center of the School of Software (Ningbo), Zhejiang University, Ningbo, China\\
$^3$ Institute of Remote Sensing Satellite, China Academy of Space Technology, Beijing, China\\
$^4$ Zhejiang Key Laboratory of Digital-Intelligence Service Technology, Hangzhou, China
}

\begin{document}

\maketitle
\footnotetext[1]{Meng Xi is the corresponding author.} 

\begin{abstract}
Text-based speech editing aims to modify specific segments while preserving speaker identity and acoustic context. Current approaches generally involve either expensive task-specific training or adapting pre-trained Text-to-Speech (TTS) models. However, both paradigms face challenges: task-specific methods often degrade fidelity in unedited regions, whereas TTS adaptations struggle with a trade-off between editing naturalness and temporal fidelity.
To address these issues, we propose AST, an Adaptive, Seamless, and Training-free speech editing framework. Built upon pre-trained TTS, AST leverages Latent Recomposition to stitch preserved source segments with synthesized targets, guaranteeing fidelity in unedited regions. To break the quality-controllability trade-off, we introduce Adaptive Weak Fact Guidance (AWFG), which modulates a mel-space signal to ensure seamless boundary transitions without disrupting the generative manifold. Furthermore, to address evaluation gaps in temporal fidelity, we propose a new benchmark suite: the LibriSpeech-Edit dataset and a novel metric, Word-level Dynamic Time Warping (WDTW). Extensive experiments demonstrate that AST consistently outperforms existing task-specific and fine-tuned speech editing methods across content accuracy, perceptual quality, speaker preservation, and temporal fidelity. Remarkably, AST achieves state-of-the-art zero-shot speech editing performance without any task-specific training or paired editing data, validating the effectiveness of latent recomposition and AWFG in bridging the quality–controllability trade-off.
\end{abstract}


\section{Introduction}
\label{sec:introduction}

Text-to-speech (TTS) synthesis has achieved near-human naturalness~\cite{shen2018natural,ren2019fastspeech,kim2021conditional}, with modern autoregressive models~\cite{wang2023neural,borsos2023audiolm} demonstrating exceptional zero-shot capabilities. A compelling application of these advances is text-based speech editing, which aims to modify specific speech segments according to a textual transcription while preserving the original speaker identity, prosody, and acoustic context. Unlike TTS, speech editing requires a delicate balance between controlled generation and quality assurance, making it invaluable for post-production and dialogue correction~\cite{tan2021editspeech,wang2022campnet}.

\begin{figure}[t]
    \centering
    \subfloat[\textit{"GIVE NOT SO \textbf{EARNEST} A MIND TO THESE \textbf{MUMMERIES} CHILD"} -- Original Speech]{
    \includegraphics[width=0.8\linewidth]{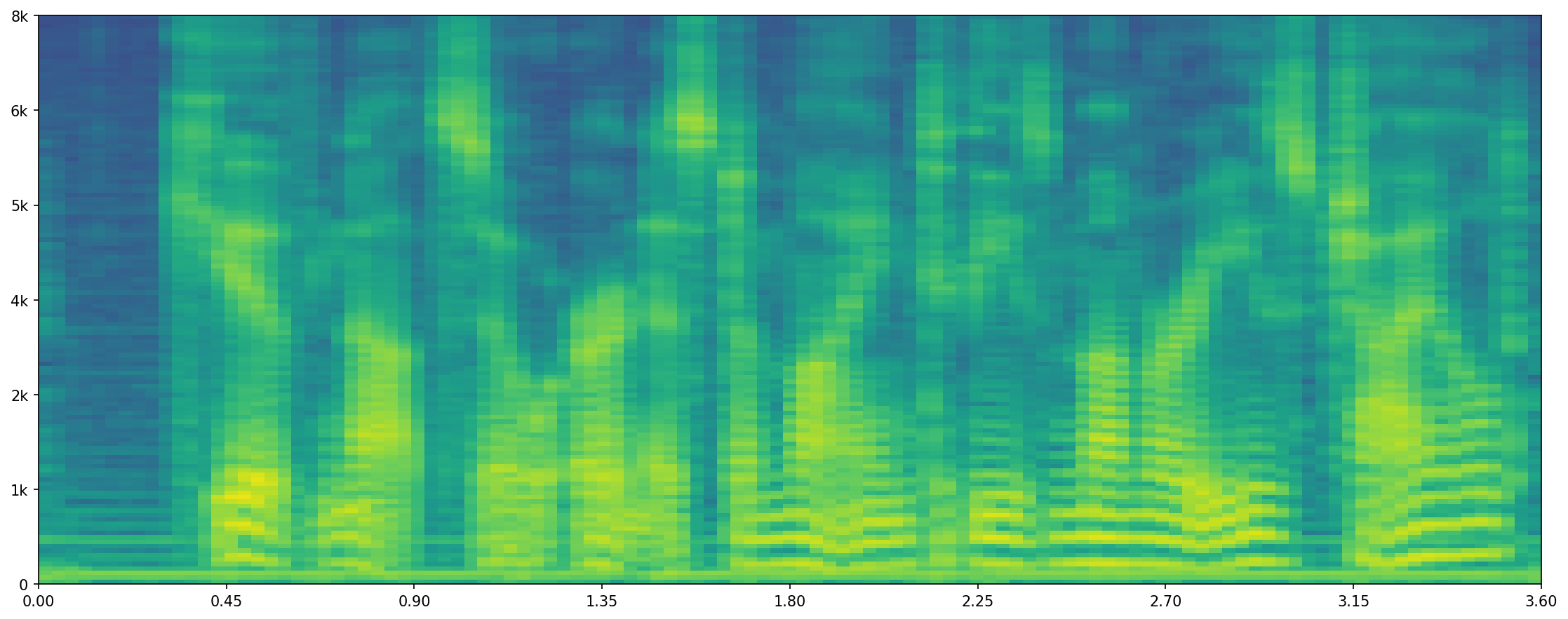}
    }
    
    \subfloat[\textit{"GIVE NOT SO \textbf{SERIOUS} A MIND TO THESE \textbf{TRICKS} CHILD"} -- Generated by TTS]{
    \includegraphics[width=0.8\linewidth]{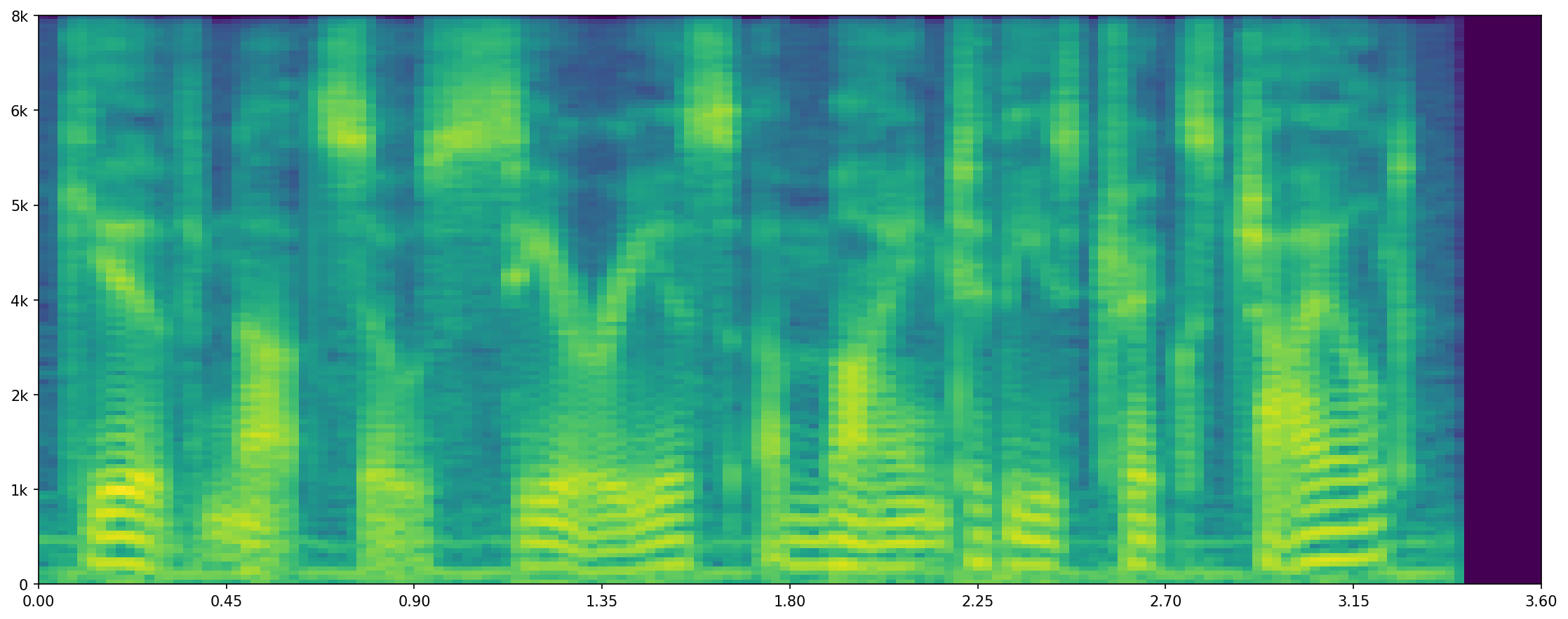}
    }
    
    \subfloat[\textit{"GIVE NOT SO \textbf{SERIOUS} A MIND TO THESE \textbf{TRICKS} CHILD"} -- Edited by AST]{
    \includegraphics[width=0.8\linewidth]{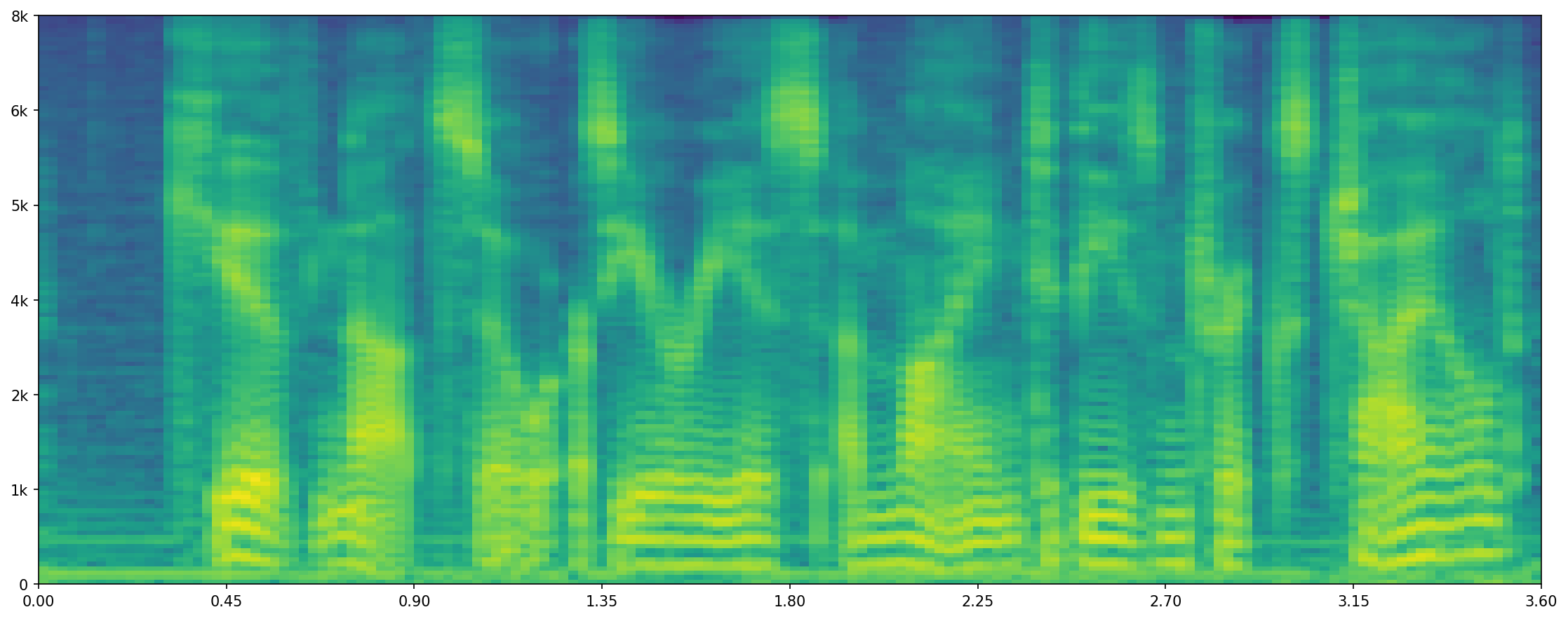}
    }
    \caption{TTS models cannot preserve the temporal characteristics of the original speech, but AST can.}
    \label{fig:rhythmic}
\end{figure}

While several methods~\cite{tan2021editspeech,wang2022campnet,jiang2023fluentspeech,peng2024voicecraft,wang2025ssr} have addressed speech editing, they typically rely on costly task-specific training and still struggle to perfectly preserve speaker identity and temporal alignment. Meanwhile, directly adapting powerful zero-shot TTS models for editing presents a fundamental challenge. Because modern autoregressive models generate entire sequences dynamically, they intrinsically reallocate global prosody and duration. Consequently, as shown in Fig.~\ref{fig:rhythmic}, naively applying them to editing inevitably leads to prosodic drifts and misaligned rhythms, destroying the original speech's structure. In image editing, latent inversion on continuous generative models~\cite{mokady2023null,huberman2024edit} successfully enables precise, zero-shot local modifications while preserving unedited regions. However, adapting these inversion-based methods to speech is fundamentally harder because the temporal dimension is inherently variable.

To overcome this quality-controllability trade-off, we propose \textbf{AST}, an \textbf{A}daptive, \textbf{S}eamless, and \textbf{T}raining-free precise speech editing framework. AST leverages the latent space structure of pre-trained flow-matching (AM-FM) TTS models. By inverting the original speech into the latent space via an inverse ODE solver, we perform Latent Recomposition: we align source and target transcripts to stitch inverted latents of unchanged regions with newly synthesized content. To prevent boundary artifacts caused by inversion approximation errors without over-constraining the generative manifold, we introduce Adaptive Weak Fact Guidance (AWFG). This mechanism dynamically modulates a mel-space guidance signal toward the original flow in preserved regions based on trajectory deviation.

To establish a sustainable benchmark, we introduce LibriSpeech-Edit, a novel dataset curated from LibriSpeech. Unlike the popular but currently inaccessible RealEdit dataset, LibriSpeech-Edit is freely available and provides a standardized evaluation protocol. Furthermore, since conventional metrics fail to measure local temporal alignment accurately, we propose Word-level Dynamic Time Warping (WDTW) to rigorously evaluate the temporal fidelity between source and edited utterances.

Extensive experiments demonstrate that AST achieves state-of-the-art controllability and temporal fidelity while maintaining highly competitive synthesis quality, all without requiring any task-specific fine-tuning. Our main contributions are summarized as follows:

\begin{itemize}
    \item We propose AST, an adaptive, seamless, and training-free speech editing framework based on latent recomposition in AM-FM paradigm TTS models, enabling precise editing while faithfully preserving speaker identity and acoustic context.
    
    \item We introduce Adaptive Weak Fact Guidance (AWFG), a novel mechanism that dynamically modulates latent trajectories to eliminate boundary artifacts without over-constraining the generative manifold.
    
    \item We release LibriSpeech-Edit, a publicly available benchmark dataset addressing previous accessibility issues, alongside Word-level Dynamic Time Warping (WDTW), a novel metric for accurately measuring local temporal alignment.
    
    \item Comprehensive experiments demonstrate that our training-free approach outperforms or matches task-specific models, establishing a new paradigm for zero-shot speech editing with state-of-the-art controllability.
\end{itemize}

\section{Related Works}
\label{sec:related_works}

Since our method addresses the speech editing task by leveraging zero-shot TTS architectures and adapting inversion techniques originally developed for visual generation, we structure our discussion into three main aspects: Neural Text-to-Speech Synthesis, Speech Editing Methods, and Latent Inversion for Generative Model Editing.

\subsection{Neural Text-to-Speech Synthesis}
\label{sec:related_tts}
Neural codec language models have revolutionized zero-shot TTS. VALL-E~\cite{wang2023neural} and its successors~\cite{borsos2023audiolm, chen2024vall} treat TTS as discrete token language modeling with strong in-context learning. Recent works further integrate flow matching into this paradigm: the CosyVoice~\cite{du2024cosyvoice,du2024cosyvoice2,du2025cosyvoice3} and IndexTTS~\cite{deng2025indextts,zhou2026indextts2} series employ Diffusion Transformer (DiT)~\cite{peebles2023scalable} backbones for high-fidelity synthesis. These AM-FM architectures provide both zero-shot generalization and a continuous latent space amenable to manipulation, forming the backbone of our editing framework.

\subsection{Speech Editing Methods}
\label{sec:related_editing}
Early approaches such as EditSpeech~\cite{tan2021editspeech} and CampNet~\cite{wang2022campnet} achieved text-based editing via partial inference or mask prediction, but relied heavily on task-specific training data. More recent methods leverage generative models: FluentSpeech~\cite{jiang2023fluentspeech} applies diffusion-based denoising for smooth partial editing, while VoiceCraft~\cite{peng2024voicecraft} and SSR-Speech~\cite{wang2025ssr} perform token infilling with codec language models. Step-Audio-Edit-X~\cite{yan2025step} further finetunes a zero-shot TTS model for joint content and emotion editing. Despite these advances, task-specific methods struggle with out-of-distribution generalization, and off-the-shelf TTS models often fail to preserve acoustic fidelity in unedited regions. Our training-free framework addresses these limitations via latent space manipulation without requiring dedicated editing data.

\subsection{Latent Inversion for Generative Model Editing}
\label{sec:related_inversion}
Latent inversion enables editing of real data without retraining by projecting signals into a model's latent space. In the image domain, DDIM inversion~\cite{song2020denoising,mokady2023null,huberman2024edit} reverses deterministic sampling to obtain latent noise representations for semantic modification. This extends naturally to flow matching~\cite{avrahami2025stable}, where the ODE formulation allows precise bidirectional integration via numerical solvers. However, latent inversion remains under-explored in speech due to unique challenges in temporal alignment. We bridge this gap by adapting inversion to AM-FM TTS models, introducing latent recomposition with adaptive guidance to achieve training-free editing that strictly preserves speaker identity and unedited acoustic context.

\begin{figure*}[t]
\centering
\includegraphics[width=0.98\linewidth]{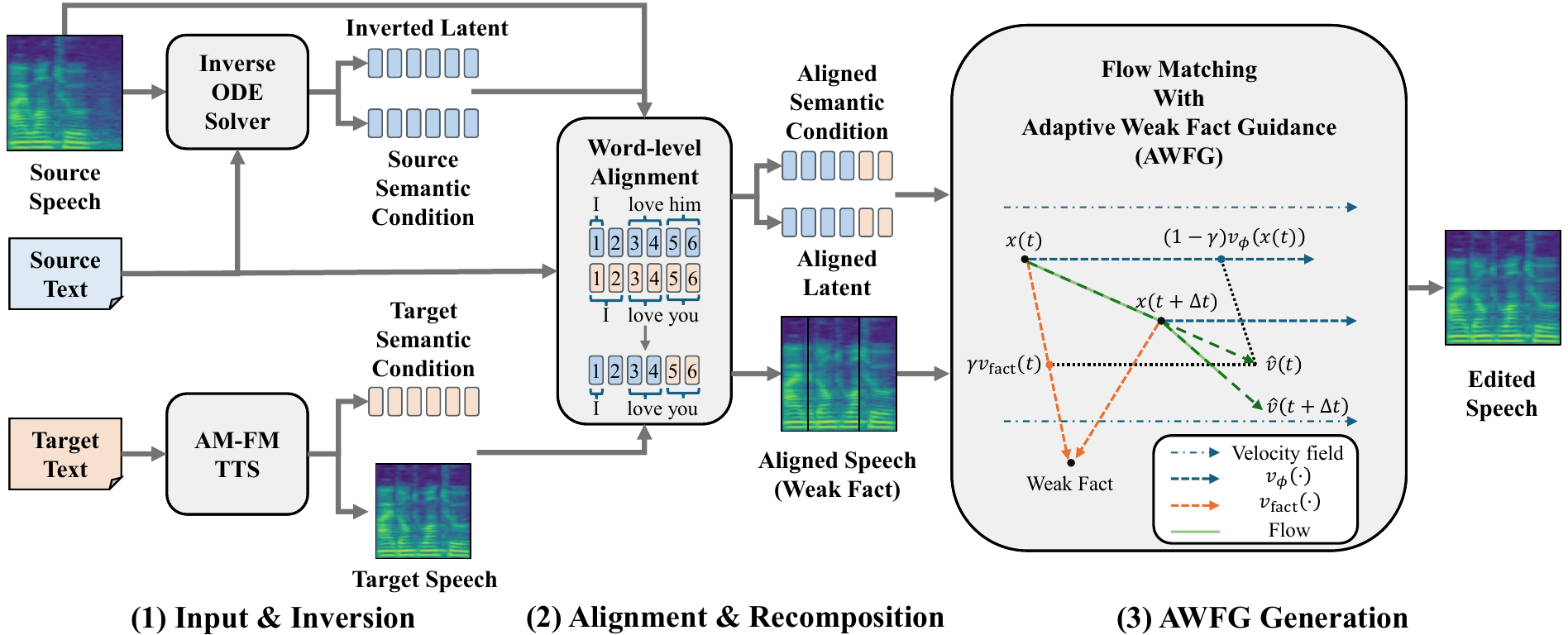}
\caption{Overview of our training-free speech editing framework, AST. The pipeline consists of three main stages: (1) \textbf{Input \& Inversion}: The source speech is inverted into the latent space via an Inverse ODE Solver, and the target text is processed by a GPT-style model to extract semantic tokens. (2) \textbf{Alignment \& Recomposition}: Source and target transcripts are aligned at the word level, recomposing inverted latent and semantic tokens, as well as constructing an aligned source speech as the weak fact. (3) \textbf{Adaptive Weak Fact Guidance (AWFG) Generation}: A Flow Matching decoder synthesizes the edited speech, utilizing AWFG to modulate the velocity field based on the recomposed latents.}
\label{fig:main}
\end{figure*}

\section{Method}
\label{sec:method}
Given a source mel-spectrogram $m_{\mathrm{ori}}$ and a target transcript $y_{\mathrm{tgt}}$, our goal is to generate an edited $m_{\mathrm{edit}}$ that reflects $y_{\mathrm{tgt}}$ while strictly preserving the original speaker identity and acoustics in unedited regions. To achieve this in a training-free manner, our AST framework (Fig.~\ref{fig:main}) leverages the semantic space of the AM-FM TTS paradigm. Rather than intervening in the continuous flow directly, we explicitly align and recompose inverted source latents with target semantics, and apply AWFG during generation to ensure stable edits and smooth transitions.

\subsection{Preliminaries}
\label{sec:prelim}

Our framework builds on the AM-FM TTS paradigm, which pairs an autoregressive semantic model with a flow matching decoder based on the DiT architecture. The FM decoder synthesizes a mel-spectrogram by integrating an ordinary differential equation (ODE):
\begin{equation}
\frac{d x(t)}{dt} = v_{\phi}\left(x(t); \mu, m_{\mathrm{ref}}\right), \quad t \in [0, 1],
\label{eqn:ode_flow}
\end{equation}
where $x(t)$ is the latent state at time $t$, $\mu$ the semantic token, and $m_{\mathrm{ref}}$ an acoustic prompt.

\subsection{Latent Inversion}
\label{sec:editing_framework}

To edit real speech, we invert the source mel-spectrogram $m_{\mathrm{ori}}$ into the latent space of the FM decoder. For a small step $\Delta t$, the forward Euler discretization of Eq.~\ref{eqn:ode_flow} is
\begin{equation}
    x(t) = x(t-\Delta t) + \Delta t \cdot v_\phi\left(x(t-\Delta t); \mu, m_{\mathrm{ref}}\right),
    \label{eq:euler-forward}
\end{equation}
Assuming the velocity field varies slowly (i.e., $v_\phi(x(t)) \approx v_\phi(x(t-\Delta t))$ for sufficiently small $\Delta t$), we reverse this step to obtain an inverse Euler solver:
\begin{equation}
    x(t-\Delta t) = x(t) - \Delta t \cdot v_\phi\left(x(t); \mu, m_{\mathrm{ref}}\right).
    \label{eq:euler-invert}
\end{equation}
By setting $x_{\mathrm{ori}}(1)=m_{\mathrm{ori}}$ and conditioning on the source semantic token $\mu_{\mathrm{ori}}$, we apply Eq.~\ref{eq:euler-invert} iteratively to recover the initial latent $x_{\mathrm{ori}}(0)$. In parallel, we generate the target mel-spectrogram $m_{\mathrm{tgt}}$ by starting from noise $x_{\mathrm{tgt}}(0)=\epsilon\sim\mathcal{N}(0,I)$, conditioning on the target semantic token $\mu_{\mathrm{tgt}}$, and integrating Eq.~\ref{eq:euler-forward} forward in time. The inverted latents and $m_{\mathrm{tgt}}$ provide the basis for the alignment and recomposition stage described next.

\subsection{Word-level Alignment and Recomposition}

Unlike image or audio editing tasks where dimensions are typically fixed, speech editing often alters the total duration and shifts specific segments. To address this structural mismatch and preserve unchanged content, we align the semantics between $y_{\mathrm{ori}}$ and $y_{\mathrm{tgt}}$. Specifically, we compute the Longest Common Subsequence of their word sequences to identify the indices of unchanged words, denoted as $\mathcal{I}_{\mathrm{match}}$. A forced alignment tool then maps each word to a contiguous interval of mel-frames. Due to natural variations in speaking rates, these intervals differ in length across words and even for the same word across utterances.

Formally, let $k$ be the index of a word in the target sequence, with $\mathcal{T}_{\mathrm{ori}}^{(k)}$ and $\mathcal{T}_{\mathrm{tgt}}^{(k)}$ denoting its corresponding mel-frame intervals in the original and target utterances. We construct the $k$-th segment of the recomposed mel-spectrogram $m_{\mathrm{fact}}^{(k)}$ via interval slicing:
\begin{equation}
    m_{\mathrm{fact}}^{(k)} = \begin{cases} m_{\mathrm{ori}}\left[\mathcal{T}_{\mathrm{ori}}^{(k)}\right], & \text{if } k \in \mathcal{I}_{\mathrm{match}} \\ m_{\mathrm{tgt}}\left[\mathcal{T}_{\mathrm{tgt}}^{(k)}\right], & \text{otherwise} \end{cases},
    \label{eq:mfact}
\end{equation}
where $[\mathcal{T}]$ extracts the sequence over interval $\mathcal{T}$. Similarly, we obtain the $k$-th segment of the recomposed semantic token $\mu_{\mathrm{fact}}^{(k)}$:
\begin{equation}
    \mu_{\mathrm{fact}}^{(k)} = \begin{cases} \mu_{\mathrm{ori}}\left[\mathcal{T}_{\mathrm{ori}}^{(k)}\right], & \text{if } k \in \mathcal{I}_{\mathrm{match}} \\ \mu_{\mathrm{tgt}}\left[\mathcal{T}_{\mathrm{tgt}}^{(k)}\right], & \text{otherwise} \end{cases},
\end{equation}
where $\mu_{\mathrm{ori}}$ and $\mu_{\mathrm{tgt}}$ are the semantic tokens of the original and target speech. The full sequences $m_{\mathrm{fact}}$ and $\mu_{\mathrm{fact}}$ are obtained by concatenating their respective segments along the time axis.

To initialize the forward Flow Matching process at $t=0$, we construct the recomposed latent variable $x_{\mathrm{edit}}(0)$ by stitching the inverted source latents with Gaussian noise:
\begin{equation}
    x_{\mathrm{edit}}^{(k)}(0) = \begin{cases} x_{\mathrm{ori}}(0)\left[\mathcal{T}_{\mathrm{ori}}^{(k)}\right], & \text{if } k \in \mathcal{I}_{\mathrm{match}} \\ \epsilon^{(k)} \sim \mathcal{N}(0, I), & \text{otherwise} \end{cases},
    \label{eqn:x_edit}
\end{equation}
where $\epsilon^{(k)}$ is Gaussian noise matching the temporal length of the target interval $\mathcal{T}_{\mathrm{tgt}}^{(k)}$. The concatenated initial latent $x_{\mathrm{edit}}(0)$ and condition $\mu_{\mathrm{fact}}$ are then integrated into the forward ODE (Eq.~\ref{eqn:ode_flow}) to synthesize the final edited mel-spectrogram $m_{\mathrm{edit}}$.

\begin{figure}[htbp]
    \centering
    \includegraphics[width=0.8\linewidth]{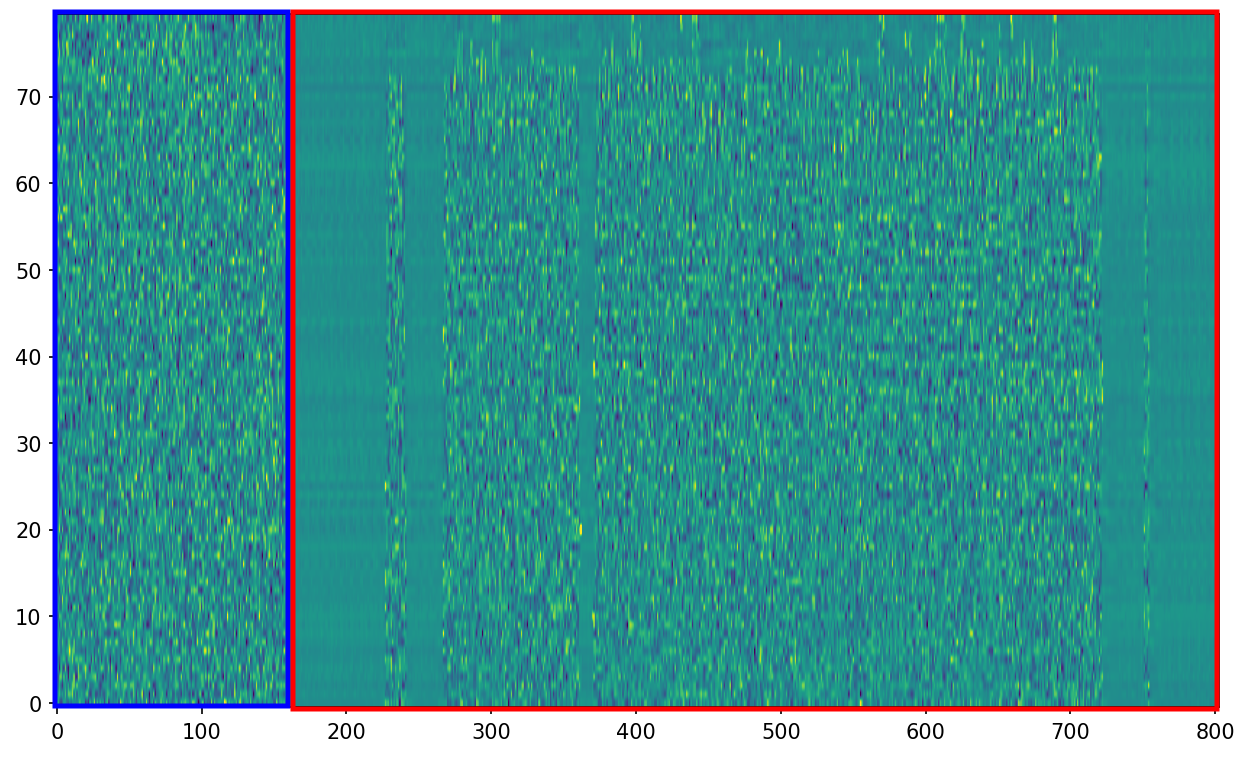}
    \caption{Illustration of the latent recomposition strategy for speech editing. The final latent sequence is constructed by stitching the inverted source latents in \textcolor{red}{matched regions} with the Gaussian noise in \textcolor{blue}{edited regions} based on text alignment.}
    \label{fig:alignment_vis}
\end{figure}

Fig.~\ref{fig:alignment_vis} illustrates this recomposition strategy. For matched regions, copying inverted latents from $x_{\mathrm{ori}}$ and conditions from $\mu_{\mathrm{ori}}$ preserves the original acoustic dynamics. Conversely, for edited regions, initializing with noise and using $\mu_{\mathrm{tgt}}$ enables new content synthesis. This latent stitching in the flow space effectively balances source prosody preservation with textual edits.

\subsection{Adaptive Weak Fact Guidance}
\label{sec:adaptive_guidance}

\begin{figure}[htbp]
    \centering
    \subfloat[Without AWFG: visible \textcolor{blue}{artifacts} at \textcolor{red}{boundaries}.]{
    \includegraphics[width=\linewidth]{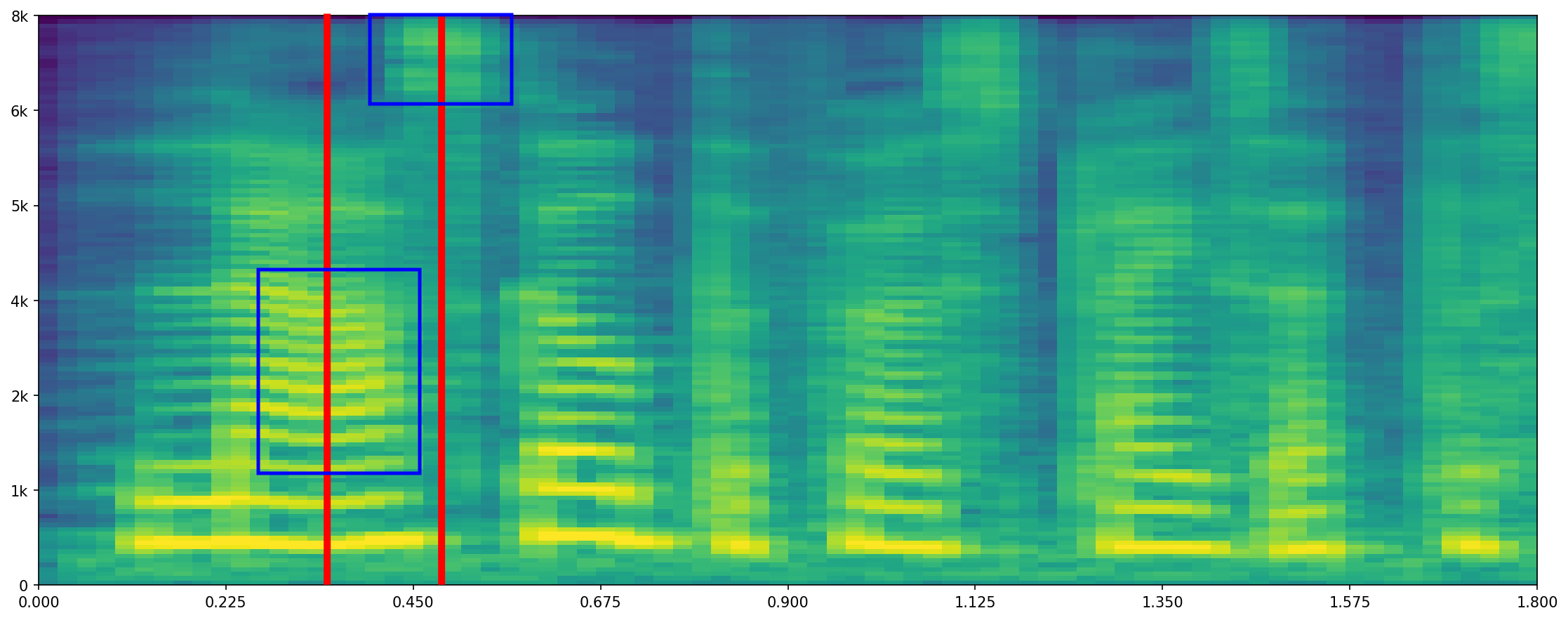}
    \label{fig:adaptive_guidance:wo}
    }
    
    \subfloat[With AWFG: smooth transitions and precise synthesis.]{
    \includegraphics[width=\linewidth]{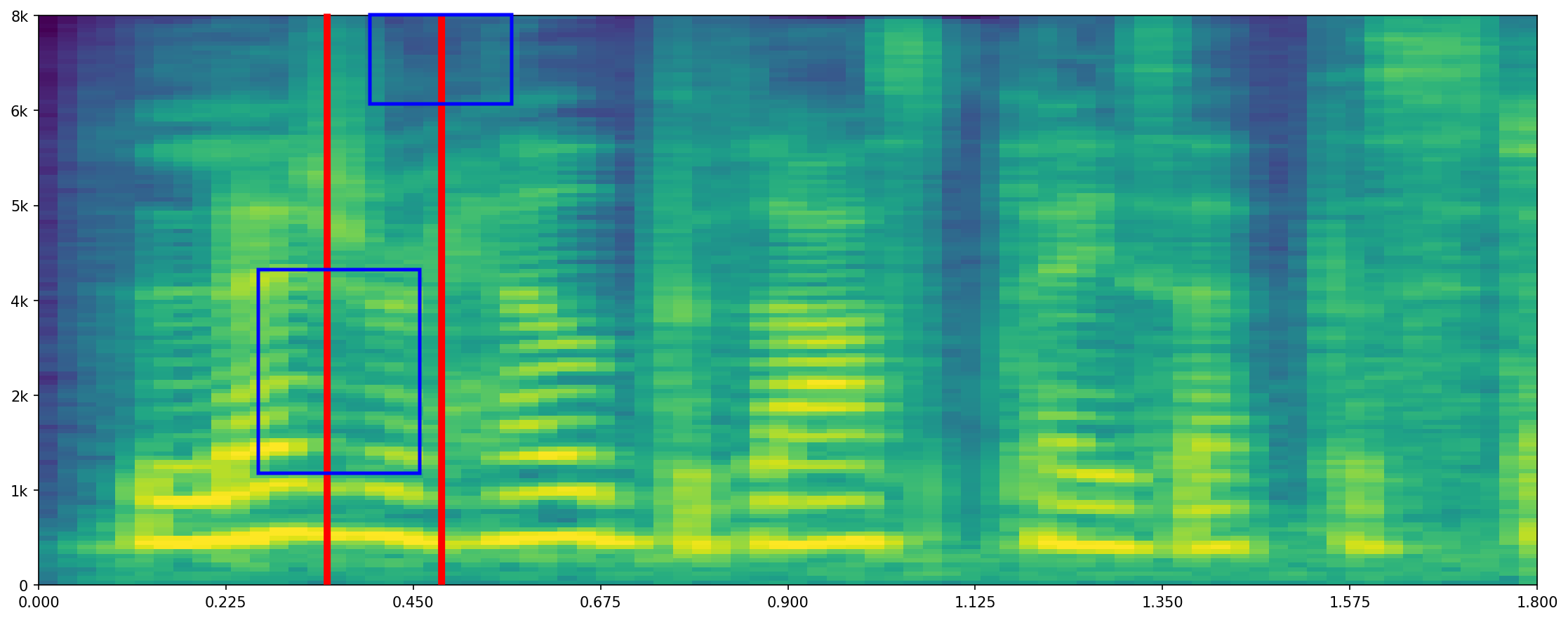}
    \label{fig:adaptive_guidance:w}
    }
    \caption{Qualitative comparison of speech synthesis at \textcolor{red}{edit boundaries}. Simple lantent alignment and recomposition suffers from \textcolor{blue}{artifacts}, while our proposed AWFG ensures seamless stitching.}
    \label{fig:adaptive_guidance}
\end{figure}

However, as Fig.~\ref{fig:adaptive_guidance:wo} demonstrates, the standard approach proves insufficient for complex edits, resulting in artifacts at stitching boundaries and unintended prosodic shifts. We hypothesize that the assumption of $v_\phi\left(x(t)\right) \approx v_\phi\left(x(t-\Delta t)\right)$ breaks down near edit boundaries, causing the learned neural dynamics to aggressively and inconsistently alter the audio during the forward process. To address this, we introduce \emph{Adaptive Weak Fact Guidance (AWFG)}: a mechanism that injects a mel-space guide signal into the ODE solving process, correcting deviations without over-constraining the generative manifold.

To anchor the generation towards the factual target, we first construct a deterministic, fact-directed velocity field $v_{\mathrm{fact}}$ with the weak fact $m_{\mathrm{fact}}$. Inspired by the Rectified Flow~\cite{liu2022flow}, we define:
\begin{equation}
v_{\mathrm{fact}}(t) = \frac{m_{\mathrm{fact}} - x(t)}{1-t},
\end{equation}
where $m_{\mathrm{fact}}$ is the recomposed mel-spectrogram in Eq.~\ref{eq:mfact}. This formulation provides a shortest-path deterministic pull towards the anchor $m_{\mathrm{fact}}$ at $t=1$, ensuring the trajectory naturally converges to the factual region without requiring additional network evaluations.

Crucially, rather than applying $v_{\mathrm{fact}}$ uniformly, we modulate its influence based on the local deviation of the current state $x$ from the desired fact flow. We measure this frame-wise discrepancy $d_k$ using the squared $L_2$ norm along the mel-channel dimension:
% 建议在导言区定义范数命令，避免在正文中重复定义
\newcommand{\norm}[1]{\left\| #1 \right\|}
\begin{equation}
d_k = \norm{ x_{\mathrm{ori}}(t)\bigl[\mathcal{T}_{\mathrm{ori}}^{(k)}\bigr] - x(t)\bigl[\mathcal{T}_{\mathrm{ori}}^{(k)}+b^{(k)}\bigr] }_c^2.
\end{equation}
where $b^{(k)}$ denotes the temporal bias introduced by recomposition. Evaluating the distance purely in the mel-channel dimension per frame captures localized spectral discrepancies while remaining invariant to global phase or amplitude shifts, making it highly robust for audio generation.

Based on $d_k$, we compute a frame-wise adaptive weight $\gamma$:
\begin{equation}
\gamma(t)[k] =
\begin{cases}
    \lambda \bigl( \mathbf{1} - e^{-d_k} \bigr), & \text{if } k \in \mathcal{I}_{\mathrm{match}} \\
    \mathbf{0}, & \text{otherwise}
\end{cases}
\label{eqn:adaptive_weight}
\end{equation}
where $\lambda \in [0, 1]$ controls the maximum guidance strength.

We arrive at the exponential form by asking what properties a gating function $g:[0,\infty)\!\to\![0,1]$, $\gamma=\lambda g(d)$, must satisfy:
\begin{enumerate}
    \item \textbf{Boundedness:} $g\le 1$ and $\lambda\le 1$ keep $\gamma\in[0,1]$, so Eq.~\ref{eqn:final_velocity} is a genuine convex combination and the guided field can never extrapolate outside the span of the two constituent fields.
    \item \textbf{Monotonicity:} $g$ increasing in $d$ closes a negative feedback loop: larger divergence elicits stronger correction, which makes tracking self-stabilizing rather than requiring a schedule tuned per utterance.
    \item \textbf{Smoothness:} $g$ should be $C^\infty$ and Lipschitz so that the guided velocity remains Lipschitz in $x$, the ODE stays well-posed, and higher-order solvers retain their order of accuracy. A hard threshold violates this and reintroduces exactly the kind of boundary discontinuity that AWFG is meant to remove.
\end{enumerate}

%The choice of the exponential mapping $1 - e^{-d_k}$ is theoretically motivated by the stability requirements of continuous ODE solvers:
%\begin{enumerate}
%    \item \textbf{Asymptotic Saturation:} When the state significantly deviates from the fact flow ($d_k \to \infty$), $\gamma$ smoothly saturates to $\lambda$. This provides a strong, uniform corrective force to pull the trajectory back, while the upper bound $\lambda$ prevents explosive gradients and over-steering that could destabilize the solver.
%    \item \textbf{Local Harmonic Approximation:} Near the target manifold where deviations are minor ($d_k \to 0$), the Taylor expansion yields $\gamma \approx \lambda d_k$. This creates a linear restoring force analogous to a harmonic oscillator, allowing the learned neural dynamics to dominate smoothly and preventing rigid artifacts.
%    \item \textbf{Lipschitz Continuity:} Unlike hard thresholding, the exponential form is infinitely differentiable. This guarantees the Lipschitz continuity of the composite vector field, preserving integral invariance and ensuring numerically stable integration.
%\end{enumerate}

Finally, the composite velocity $\tilde{v}$ is obtained via convex mixing:
\begin{equation}
\tilde{v}(t) = \left(1-\gamma(t)\right) \odot v_\phi\left(x(t); \mu_{\mathrm{fact}}, m_{\mathrm{ref}}\right) + \gamma(t) \odot v_{\mathrm{fact}}(t),
\label{eqn:final_velocity}
\end{equation}
where $\odot$ denotes element-wise multiplication along the frame sequence. Convex mixing guarantees that the modified vector field remains a valid tangent vector within the generative manifold, strictly bounding the velocity magnitude and preserving the diffeomorphic properties of the original ODE flow. Within the non-matching editing regions, $\gamma=0$ ensures content is generated with complete freedom. As Fig.~\ref{fig:adaptive_guidance} shows, this modification significantly reduces reconstruction errors and constrains edits to the intended regions.

\section{Experiments}
\label{sec:experiments}

In this section, we detail how we ran our experiments and present our results compared with existing work. We also measured how $\lambda$ and AWFG affect AST's performance.
\subsection{Experimental Setup}
\label{sec:setup}

\subsubsection{Dataset}
\label{sec:dataset}

To address the copyright-induced unavailability of the widely used RealEdit dataset~\cite{peng2024voicecraft}, we introduce \textbf{LibriSpeech-Edit}, an open and reproducible benchmark curated from the \textit{test-clean} and \textit{test-other} subsets of LibriSpeech~\cite{panayotov2015librispeech}. We employ Qwen3.6-27B~\cite{qwen3.6-27b} to generate target transcripts for the source utterances, resulting in 5,559 high-fidelity editing pairs. Figure~\ref{fig:data_statistics} summarizes the dataset statistics.

\begin{figure}[t]
\centering
\includesvg[width=\linewidth]{figs/edit_comparison_clean_vs_other.svg}
\caption{Statistics of the LibriSpeech-Edit dataset.}
\label{fig:data_statistics}
\end{figure}

\subsubsection{Baselines}
\label{sec:baselines}

We evaluate AST against four representative baselines across different speech editing paradigms:
\begin{itemize}
    \item  \textbf{FluentSpeech}~\cite{jiang2023fluentspeech}: A task-specific, diffusion-based denoising speech editing model.
    \item \textbf{SSR-Speech}~\cite{wang2025ssr}: A task-specific, codec-based speech editing model.
    \item \textbf{Step-Audio-EditX}~\cite{yan2025step}: An autoregressive TTS model fine-tuned specifically for speech editing.
    \item \textbf{IndexTTS-2}~\cite{zhou2026indextts2}: The unmodified, pretrained backbone of our framework, serving as a zero-shot baseline to evaluate our training-free improvements.
\end{itemize}

\subsubsection{Evaluation Metrics}
\label{sec:metrics}

We assess the edited speech using one subjective metric and four objective metrics:
\begin{itemize}
    \item \textbf{Mean Opinion Score (MOS)}: A subjective quality metric defined as the average rating given by 10 listeners on 50 sampled audios, following ITU-T P.800~\cite{rec1996p}.
    \item \textbf{Word Error Rate (WER)}: Measures textual accuracy and intelligibility.
    \item \textbf{DNSMOS}~\cite{reddy2021dnsmos}: A neural network-based proxy for MOS to assess overall quality.
    \item \textbf{Speaker Similarity (SpkSim)}: Computes the cosine similarity of WavLM~\cite{chen2022wavlm} speaker embeddings to measure identity preservation.
    \item \textbf{Word-level Dynamic Time Warping (WDTW)}: Our proposed metric for evaluating the temporal fidelity of edited utterances. Unlike utterance-level metrics, WDTW aligns words between the source and edited utterances to evaluate local temporal fidelity. The detailed procedure, involving forced alignment, segment-level DTW computation, and length normalization, is outlined in Algorithm~\ref{alg:wdtw}.
\end{itemize}

\begin{algorithm}[htbp]
\caption{Word-level Dynamic Time Warping (WDTW)}
\label{alg:wdtw}
\begin{algorithmic}[1]
\REQUIRE Original speech $s_{\mathrm{ori}}$, edited speech $s_{\mathrm{edit}}$, original transcript $y_{\mathrm{ori}}$, target transcript $y_{\mathrm{tgt}}$
\ENSURE WDTW score $D_{\mathrm{WDTW}}$
\STATE \textbf{// Stage 1: Forced Alignment}
\STATE $\mathcal{T}_{\mathrm{ori}} \gets \text{ForceAlign}(s_{\mathrm{ori}}, y_{\mathrm{ori}})$ 
\STATE $\mathcal{T}_{\mathrm{edit}} \gets \text{ForceAlign}(s_{\mathrm{edit}}, y_{\mathrm{tgt}})$ 
\STATE \textbf{// Stage 2: Segment Extraction \& DTW}
\STATE $L_{\mathrm{total}} \gets 0, Seg_{\mathrm{ori}}\gets [], Seg_{\mathrm{edit}} \gets []$
\FOR{each word $w_i \in y_{\mathrm{ori}}$}
    \STATE Extract duration: $Seg_{\mathrm{ori}}^{(i)} \gets (w_i, \text{Duration}(\mathcal{T}_{\mathrm{ori}}[w_i]))$
    \STATE $L_{\mathrm{total}} \gets L_{\mathrm{total}} + \text{Duration}(\mathcal{T}_{\mathrm{ori}}[w_i])$
\ENDFOR
\FOR{each word $w_i \in y_{\mathrm{tgt}}$}
    \STATE Extract duration: $Seg_{\mathrm{edit}}^{(i)} \gets (w_i, \text{Duration}(\mathcal{T}_{\mathrm{edit}}[w_i]))$
    \STATE $L_{\mathrm{total}} \gets L_{\mathrm{total}} + \text{Duration}(\mathcal{T}_{\mathrm{edit}}[w_i])$ % Fixed typo here
\ENDFOR
\STATE $D_{\mathrm{WDTW}} \gets \text{DTW}(Seg_{\mathrm{ori}}, Seg_{\mathrm{edit}})$
\STATE \textbf{// Stage 3: Normalization}
\STATE $D_{\mathrm{WDTW}} \gets D_{\mathrm{WDTW}} / L_{\mathrm{total}}$
\RETURN $D_{\mathrm{WDTW}}$
\end{algorithmic}
\end{algorithm}

\subsubsection{Implementation Details}
\label{sec:implementation}

Experiments are conducted on a single NVIDIA RTX 5880 Ada GPU using IndexTTS-2 as the backbone. The guidance strength $\lambda$ is set to $0.8$ by default, which is selected by experiments on \textit{dev-clean} and \textit{dev-other} subset. For metric computation, we use Qwen3-ASR~\cite{shi2026qwen3} for transcription and Qwen3-ForcedAligner-0.6B~\cite{shi2026qwen3} for word-level forced alignment. Since each configuration is run only once, we report standard deviations using bootstrapping.

\subsection{Main Results}
\label{sec:main_results}

\begin{table*}[t]
\caption{Experimental results on the LibriSpeech-Edit dataset. The best results are in bold, the second-best underlined.}% 
\centering
\small
\begin{threeparttable}
\begin{tabular}{l l c c c c c}
\toprule
\textbf{Method} & \textbf{Approach} &  \textbf{MOS $\uparrow$} & \textbf{WER $\downarrow$} & \textbf{DNSMOS $\uparrow$} & \textbf{SpkSim $\uparrow$} & \textbf{WDTW $\downarrow$} \\
\midrule
%\multicolumn{6}{c}
\textbf{test-clean}&&&&& \\
\midrule
FluentSpeech & task-specific model & 3.356 & 0.2593 $\pm$ .0052 & 3.448 $\pm$ .009 & 0.9455 $\pm$ .0010 & 0.3266 $\pm$ .0045 \\
SSR-Speech & task-specific model & \underline{3.986} & \underline{0.0508 $\pm$ .0025} & \underline{3.812 $\pm$ .006} & \underline{0.9732 $\pm$ .0012} & \underline{0.2189 $\pm$ .0032} \\
Step-Audio-EditX & fine-tuned TTS & 3.844 & 0.1327 $\pm$ .0028 & 3.754 $\pm$ .006 & 0.9573 $\pm$ .0008 & 0.2363 $\pm$ .0034 \\
IndexTTS-2 & pre-trained TTS & 3.880 & 0.0565 $\pm$ .0024 & 3.738 $\pm$ .005 & 0.9678 $\pm$ .0008 & 0.2970 $\pm$ .0029 \\
AST & training-free & \textbf{4.044} & \textbf{0.0359} $\pm$ \textbf{.0017} & \textbf{3.842} $\pm$ \textbf{.005} & \textbf{0.9806} $\pm$ \textbf{.0007} & \textbf{0.2162} $\pm$ \textbf{.0030} \\
\midrule
\textbf{test-other}&&&&& \\
\midrule
FluentSpeech & task-specific model & 3.076 & 0.3113 $\pm$ .0054 & 3.239 $\pm$ .009 & 0.9289 $\pm$ .0013 & 0.3736 $\pm$ .0046 \\
SSR-Speech & task-specific model & \underline{3.754} & 0.0923 $\pm$ .0032 & 3.644 $\pm$ .006 & 0.9503 $\pm$ .0017 & \underline{0.2774 $\pm$ .0036} \\
Step-Audio-EditX & fine-tuned TTS & 3.668 & 0.1765 $\pm$ .0035 & \underline{3.650 $\pm$ .006} & 0.9419 $\pm$ .0012 & 0.2900 $\pm$ .0035 \\
IndexTTS-2 & pre-trained TTS & 3.648 & \underline{0.0885 $\pm$ .0029} & 3.566 $\pm$ .005 & \underline{0.9550 $\pm$ .0010} & 0.3420 $\pm$ .0029 \\
AST  & training-free & \textbf{3.822} & \textbf{0.0670} $\pm$ \textbf{.0022} & \textbf{3.667} $\pm$ \textbf{.005} & \textbf{0.9747} $\pm$ \textbf{.0008} & \textbf{0.2478} $\pm$ \textbf{.0032} \\
\bottomrule
\end{tabular}
\end{threeparttable}
\label{tab:main_results}
\end{table*}

Our training-free AST framework consistently outperforms all task-specific and fine-tuned baselines across every evaluation dimension on both the \textit{test-clean} and \textit{test-other} subsets, establishing a new state of the art for zero-shot speech editing. As shown in Table~\ref{tab:main_results}, AST achieves the best accuracy, quality, speaker and temporal fidelity in all settings, without any task-specific training or paired editing data. This directly validates our core claim that latent inversion and recomposition within pretrained AM-FM TTS models can fully bridge the quality–controllability trade-off.

AST delivers the most accurate content editing while producing the most natural-sounding speech. On \textit{test-clean}, it reduces the WER of the strongest task-specific baseline, SSR-Speech, from 0.0508 to 0.0359, and pushes DNSMOS to 3.842, surpassing all compared systems. Correspondingly, the subjective MOS score of AST reaches 4.044, significantly exceeding SSR-Speech and Step-Audio-EditX. This consistent superiority in both objective and subjective quality stems from our latent recomposition and AWFG, which effectively suppress boundary artifacts and stabilize generation near edit regions. In contrast, the fine-tuned Step-Audio-EditX suffers from a substantially higher WER of 0.1327, indicating that naive TTS adaptation is insufficient for maintaining textual fidelity.

Speaker identity is preserved with exceptional fidelity. AST attains the highest SpkSim scores of 0.9806 and 0.9747 on \textit{test-clean} and \textit{test-other}, outperforming both SSR-Speech and IndexTTS-2 by a clear margin. This is because the flow trajectories for unmodified segments are initialized from exactly inverted latent states of the source speech rather than being regenerated from scratch. The purely generative IndexTTS-2 baseline achieves a lower SpkSim of 0.9678 on \textit{test-clean}, highlighting the difficulty of reconstructing fine speaker characteristics without explicit preservation.

Temporal fidelity, measured by our proposed WDTW, is a critical dimension of speech editing that conventional metrics miss. AST records the lowest WDTW values in both conditions, significantly outperforming the pre-trained IndexTTS-2 baseline. This confirms that our inverted latent recomposition rigorously constrains unedited regions to their original rhythm, whereas directly using the TTS model inevitably reassigns global prosody and causes temporal drift.

AST also exhibits strong robustness under acoustic domain shift, maintaining its leadership across all metrics on the more challenging \textit{test-other} subset with a WER of 0.0670 and a SpkSim of 0.9747. We attribute this advantage to the strict preservation of unedited regions rather than to any learned invariance: by directly initializing unchanged segments from the exactly inverted source latents, AST faithfully reproduces the original acoustic characteristics regardless of recording conditions. In contrast, task-specific baselines must regenerate these regions from scratch, and the uncertainty introduced by this process is further amplified under domain mismatch.

\subsection{Hyperparameter Analysis}
\label{sec:hyperparameter}

\begin{figure}[htbp]
\centering
\includesvg[width=\linewidth]{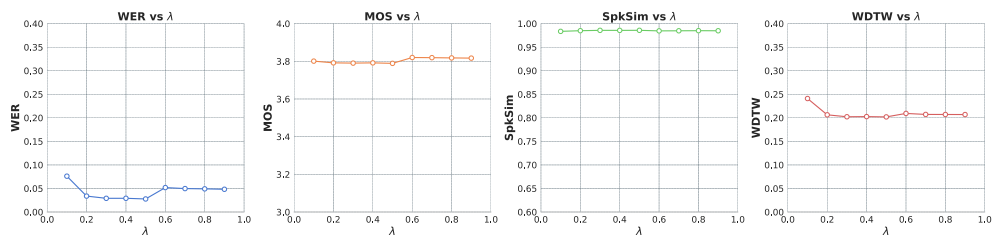}
\caption{Performance variation with respect to the maximum guidance strength $\lambda$.}
\label{fig:hyperparameter}
\end{figure}

The maximum guidance strength \(\lambda\) governs the influence of the fact-directed velocity field in AWFG, effectively controlling the trade‑off between strict temporal/spectral alignment and the creative freedom of the learned generative prior. Figure~\ref{fig:hyperparameter} tracks the evolution of all metrics as \(\lambda\) varies from \(0.1\) to \(0.9\). Overall, AST proves highly robust to this choice: every indicator remains at a competitive level across the full range, and no single \(\lambda\) dominates in all dimensions simultaneously.

A clear monotonic improvement in editing precision and temporal fidelity emerges as \(\lambda\) increases from weak values towards \(0.8\). Both WER and WDTW consistently decrease, with the most pronounced gains occurring between \(\lambda = 0.5\) and \(0.8\); beyond this point the curves flatten, indicating that the guidance has saturated its corrective capacity. This behaviour directly validates the AWFG mechanism: stronger fact guidance more effectively suppresses boundary artifacts and enforces the original temporal structure of unedited words, while excessively large \(\lambda\) offers no extra benefit and merely reduces the solver’s flexibility. Speaker similarity exhibits a mild but steady upward tendency over the same interval, suggesting that anchoring the flow to the recomposed fact also helps preserve fine speaker characteristics by inhibiting spurious variations.

In contrast, the DNSMOS curve shows a slight initial dip at intermediate \(\lambda\) values, followed by a gentle recovery as \(\lambda\) grows further. The overall fluctuation is extremely small, and the lowest point remains substantially higher than the DNSMOS scores of the best task‑specific baselines reported in Table~\ref{tab:main_results}. This non‑monotonic pattern reflects the dual role of guidance: when too weak, it cannot fully eliminate stitching artifacts, which slightly degrade perceived naturalness; as it becomes effective, those artifacts are removed and quality is restored. The minute magnitude of the variation confirms that the generative manifold is never over‑constrained, exactly as intended by the adaptive exponential mapping of AWFG. 

\subsection{Ablation Study}
\begin{figure}[htbp]
    \centering
    \includesvg[width=\linewidth]{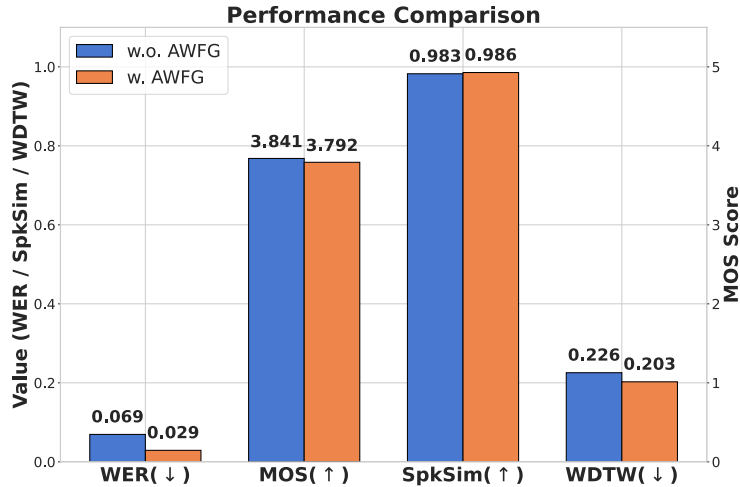}
\caption{Ablation study on the effectiveness of AWFG.}
    \label{fig:ablation}
\end{figure}
To isolate the contribution of AWFG, we compare the full AST against a variant that drops the guidance entirely while keeping the latent recomposition pipeline unchanged. The quantitative comparison in Fig.~\ref{fig:ablation} reveals that AWFG is indispensable for precise and temporally faithful editing. On the \textit{test‑clean} subset, removing AWFG causes the WER to rise from \(0.0359\) to \(0.0563\) and WDTW to degrade from \(0.2162\) to \(0.2346\); analogous relative deteriorations appear on \textit{test‑other}. These numbers quantitatively confirm the earlier qualitative observation: merely stitching inverted latents without boundary correction leaves the flow dynamics free to introduce errors at edit junctions. Speaker similarity also profits from the guidance, showing a clear increase because the fact‑anchored flow better retains the original acoustic fingerprint in unchanged segments.

As expected, DNSMOS undergoes a very minor reduction when AWFG is active, decreasing by only \(0.019\) on \textit{test‑clean}. This drop is imperceptible in practice and, critically, even the guided AST still outperforms baselines by a comfortable margin. The small quality-controllability trade‑off is intrinsic to any guidance mechanism: softly steering the trajectory towards the fact flow slightly limits the model’s generative freedom, but the sacrifice in naturalness is negligible compared to the substantial gains in edit accuracy, temporal alignment, and speaker fidelity. 

Overall, the hyperparameter analysis and ablation study jointly demonstrate that AWFG is a well‑calibrated, highly effective component. It eliminates boundary artifacts and enforces local temporal consistency with virtually no impact on synthesis quality, enabling AST to realise the full potential of training‑free speech editing by maintaining an optimal balance between fidelity and naturalness.

\section{Conclusion}

In this paper, we have presented AST, an adaptive, seamless, and training‑free speech editing framework that leverages latent inversion and recomposition within pretrained AM‑FM TTS models. By inverting source speech into the latent space and stitching unchanged regions with newly synthesized content, we have enabled precise local edits without any task‑specific fine‑tuning. To eliminate boundary artifacts caused by inversion approximation errors, we have introduced AWFG, which dynamically modulates a mel‑space guidance signal to preserve temporal structure while avoiding over‑constraint. To establish a reproducible evaluation standard, we have released LibriSpeech‑Edit, a publicly available benchmark dataset, and proposed WDTW to rigorously assess local temporal fidelity. Extensive experiments on LibriSpeech‑Edit have demonstrated that AST consistently achieves state‑of‑the‑art controllability and temporal fidelity, outperforming strong task‑specific baselines in WER, DNSMOS, speaker similarity, and WDTW, all without requiring any dedicated editing data or training. Our training‑free paradigm has shown that the rich generative priors of zero‑shot TTS models can be effectively harnessed for high‑quality speech editing. We believe it opens new directions for faithful speech manipulation.

\section*{Acknowledgments}
This work is supported by the National Natural Science Foundation of China (62502427), the Science and Technology Program of Zhejiang Province (2025C01087), the Yongjiang Talent Introduction Program (2024A-404-G), the Major Scientific and Technological Projects of CNTC(110202401031(SZ-05)), and the Zhejiang Key Laboratory Project (2024E10001).

\bibliographystyle{unsrt}
\bibliography{base}

\end{document}